\documentclass[twocolumn]{revtex4} 

\usepackage[utf8]{inputenc}
\usepackage{xcolor}
\usepackage[utf8]{inputenc}
\usepackage{graphicx}
\usepackage{amssymb}
\usepackage{amsmath}
\usepackage{xcolor}
\usepackage{amsmath,amssymb} 
\usepackage{epsfig}   
\usepackage[english]{babel}

\newcommand{\new}{\textcolor{black}}

\begin{document}
\title{Influence of density-dependent diffusion on pattern formation in a refuge}

\author{G. G. Piva$^{1}$, 
C. Anteneodo$^{1,2}$}
\address{$^{1}$Department of Physics, Pontifical Catholic University of Rio de Janeiro, PUC-Rio, and 
$^{2}$National Institute of Science and Technology for Complex Systems, INCT-CS, Rua Marqu\^es de S\~ao Vicente 225, 22451-900, Rio de Janeiro, RJ, Brazil}

\begin{abstract}

We investigate a nonlocal generalization of the Fisher-KPP equation, which incorporates logistic growth and diffusion,  for a single species population in a viable patch (refuge). 
In this framework, diffusion plays an homogenizing role, while nonlocal interactions can destabilize the spatially uniform state, leading to the emergence of spontaneous patterns. 
Notably, even when the uniform state is stable, spatial perturbations, such as the presence of a refuge,  can still induce patterns.
These phenomena are well known for environments with constant diffusivity. Our goal is to investigate how the formation of winkles in the population distribution is affected when the diffusivity is density-dependent. 
Then, we explore scenarios in which diffusivity is sensitive to either rarefaction or overcrowding. 
We find that state-dependent diffusivity affects the shape and stability of the patterns, potentially leading to either explosive growth or fragmentation of the population distribution, depending on how diffusion reacts to changes in density.
\end{abstract}


\maketitle

\section{Introduction}
\label{sec:intro}

  A remarkable property of biological systems  is the formation of spatial structures.  
Patterns can emerge by self-organization as a result of specific interactions between individuals, without the need for external drivers~\cite{von2001diversity, rietkerk2008regular, rietkerk2002self}. 
In particular, in the framework of Fisher-type dynamics, which includes logistic growth and diffusion~\cite{murray2002},  when competitive interactions are spatially extended (nonlocal), they can give rise to self-organized spatial oscillations, with the characteristic wavelength determined by the range of these interactions~\cite{fuentes2003nonlocal,fuentes2004analytical, hernandez2004clustering}. 
 For instance, plant competition for water, known for generating spatial patterns, can be considered a nonlocal process due to root spatial structure or water diffusive dynamics~\new{\cite{escaff2015localized,martinez2013vegetation}.} The nonlocality of other elementary processes, such as reproduction and random dispersal, has a less central role but can interfere constructively or destructively with the possibility of pattern formation~\cite{piva2021interplay}. 
 Furthermore, while within Fisher dynamics diffusion can be detrimental to pattern formation, since it promotes homogenization, it can still influence the shape and stability of the resulting patterns, depending on the kind of diffusion (e.g., normal or anomalous)~\cite{colombo2012nonlinear}.

Although patterns can emerge solely from interactions among individuals, 
  they are naturally affected or even induced by environmental conditions, as these control the rates of biological processes~\cite{smith2003animal,corley1977influence}. 
\new{Such effects can be observed at ecological scales in nature, as in the case of vegetation patterns induced by spatial heterogeneities~\cite{sheffer2012,tarnita2017,pintoramos2022},} as well as artificially produced in the laboratory, such as when bacterial colonies are subjected to adverse conditions such as ultraviolet light, except for a protected area (refuge)~\cite{perry2005experimental}. 
In fact, growth-rate heterogeneity can induce pattern formation, even under conditions that will not give rise to patterns in homogeneous media~\cite{dornelas2021landscape}. 

Furthermore,  in real environments, not only growth rates, as in the case of a refuge, but also mobility can be heterogeneous, e.g., in active suspensions where movement is influenced by nutrient gradients~\cite{diluzio2005escherichia}, or due to structural features of the environment, as the movement of bacteria in porous media~ \cite{sosa2017motility,muskat1937flow}. 
Furthermore, heterogeneous diffusion can also emerge as a reaction~\cite{arlt2019dynamics} that the individuals manifest, for instance, in response to overcrowding or sparsity
 of a population, favoring, or not, the random motion among other
individuals~\cite{colombo2012nonlinear,matthysen2005, kun2006, french2001, Cates2010,murray2002, lopez2006macroscopic, kenkre2008nonlinearity, lenzi2001escape,anteneodo2005non,anteneodo2007brownian,newman1980some,gurtin1977diffusion, kareiva1983local, birzu2019genetic}. 
The specific way organisms respond to concentration depends on several conditions and vary from species to species~\cite{gurtin1977diffusion,kareiva1983local,newman1980some,murray2002}. 
For instance, in populations of grasshoppers, the diffusion coefficient is enhanced at high densities, where encounters between individuals are more frequent, but in other species, this occurs at low densities~\cite{murray2002}. 
Another source of heterogeneous diffusion, accompanied by bias, is chemotaxis~\cite{mittal2003motility, hillen2009user,arumugam2021keller, kenkrel2008nonlinearity}. 
Density-dependent factors are also present in migratory dispersal~\cite{amarasekare2004role,Matthysen2005density,kun2006}.
State-dependent diffusivity in biological population dynamics has been previously considered in the context of critical conditions for survival~\cite{amarasekare2004role, colombo2012nonlinear, maciel2013individual}, \new{and also with regard to pattern formation  
\cite{liu2013,liu2016}. }
Let us mention an important study on the distinct roles on pattern formation of Fokker-Planck and Fick's laws of diffusion, for spatially varying coefficient of diffusion, $D(x)$~\cite{bengfort2016fokker}. 
Although we will  address systems with nonlocality as pattern-formation mechanism, let us mention that there are works showing that the  effects of nonhomogeneous environments cannot be neglected in systems with Turing instabilities~\cite{Yizhaq2014,Cobbold2015}.

In this work, we analyze two classes of heterogeneous diffusivity: state-dependent (where the diffusivity responds to the population density) and space-dependent (where diffusivity is associated to the quality of the environment). In both cases there might be a feedback that mitigates or reinforces pattern formation. 
Moreover, we focus on the effects that diffusive heterogeneities have on the spatial distribution of a population 
inhabiting a refuge immersed in an adverse environment. 
We consider as starting point the description of a single-population dynamics given by  the spatially-exteded (nonlocal) form of the Fisher-Kolmogorov-Petrovsky-Piskunov (FKPP) dynamics~\cite{murray2002}, which   includes  random movements (normal diffusion) and logistic growth with  nonlocal competition. Then we generalize this equation by substituting  normal diffusion by each form of heterogeneous diffusion considered.
For state-dependent diffusivity, we focus on two functional  forms, namely, decay or increase with the population density, reflecting enhanced mobility in response to sparseness and overcrowding, respectively. For this purpose, exponential dependencies are studied as paradigm. 
For the spatial-dependent case, we consider that diffusivity is associated to the quality of the environment, which is different inside and outside the refuge.   
 
By numerical integration of the effective dynamics equation, we find that heterogeneous diffusivity does not affect significantly the critical conditions for pattern formation
within the refuge, in agreement with the theoretical linear-stability analysis, but heterogeneity does affect the shape and stability of the patterns.
This study may bring insights, for instance, on observations made in experiments with bacteria~\cite{perry2005experimental},   where puzzling results cannot be explained by considering  only the simple form of the FKPP equation.

\section{Model}

We consider a single-species population living in a focal patch  of size $L$ immersed in a large hostile environment in one dimension, scenario mimicked by a positive
growth rate $r_{in}>0$ inside the refuge, and a negative one $r_{out}<0$ outside, namely
\begin{equation} \label{eq:rx}
    r(x) = r_{in}  + (r_{out}-r_{in}) \Theta(|x|-L/2)\,, 
\end{equation}
being $\Theta$ the Heaviside step function. 
 
Then, the generalized FKPP dynamics, with heterogeneity and spatially-extended competition, for the population density $u(x,t)$, becomes
\begin{equation}
\partial_t u = \partial_x (D(u,x) \partial_x u) + r(x) u -u(\gamma \star u), 
\label{eq:FKPP}
\end{equation}
where the symbol ``$\star$" stands for the convolution operation that provides nonlocality through an interaction  kernel $\gamma$, which  for simplicity we consider to be a normalized rectangular shape of width $2w$. 
This kind of system was considered before, for constant diffusion coefficient $D$, that is, for  homogeneous diffusivity~\cite{dornelas2021landscape}. 
The extension we propose below, inspired in previous literature~\cite{Cates2010,tailleur2008statistical,von2001diversity, rietkerk2008regular, rietkerk2002self}, assumes that the diffusivity  can depend on the density and/or on the spatial coordinate directly, $D(u,x)$,  
reflecting  a reaction of the  mobility in response to the distribution of other individuals or to a hostile medium.

\subsection{State-dependent diffusivity}

We are mainly interested in variations of the diffusivity that are self-generated, as response to the population level. 
First in Sec.~\ref{sec:stateD}, we will investigate a decreasing function of the population density, namely,  
\begin{equation} \label{eq:Du}
   D_1(u)=d \exp(-u/\sigma)\,, 
\end{equation} 
where $d$ and $\sigma$ are positive parameters, such that $\sigma$ controls the decay with density, recovering a homogeneous diffusivity profile in the limit  $\sigma\to\infty$. 
This choice was motivated by previous work assuming that density has a negative impact on diffusion~\cite{Cates2010,tailleur2008statistical}. 
This functional form of $D_1(u)$ reflects a reaction to sparsity, with greater mobility the more rarefied the population is. 
In our case of a refuge within a hostile environment, the functional form of $D_1$ implies a lower diffusivity inside, where the population is more dense since $r_{out}<r_{in}$.

For the opposite possibility of enhanced response to overcrowding, we will use as counterpart of Eq.~(\ref{eq:Du}), 
\begin{equation}  \label{eq:Du2}
    D_2(u)=d[(1-\exp(-u/\sigma)] ,
\end{equation} 
for which homogeneity is obtained in the opposite limit 
$\sigma \to 0$.

\subsection{Space-dependent diffusivity}
  
As another relevant  case, we will consider a diffusivity profile  $D(x)$ taking the values $D_{in}$ and $D_{out}$ inside and outside the refuge, respectively, namely
\begin{equation} \label{eq:Dx}
    D(x)=D_{out}+(D_{in}-D_{out})\Theta_s [L/2-|x|],
\end{equation}
where $\Theta_s$ is a smoothed Heaviside step function, and the jump has width $s$, such that the usual Heaviside is recovered in the limit $s\to 0$.
Similar settings have been used to study the role of space-dependent diffusion on the critical patch size~\cite{cantrell1999diffusion,maciel2013individual, dos2020critical}.  In Sec.~\ref{sec:spaceD}, we will 
investigate its impact on pattern formation.

\section{Methods}
\label{sec:methods}

 Results for the  different scenarios described above, focusing on pattern formation and mode stability, will be shown in the next sections. 
In all cases,  Eq.~(\ref{eq:FKPP}) was numerically integrated  using a forward-time centered-space algorithm (typically, $\Delta x=0.02$ and $\Delta t=10^{-5}$), with periodic boundary conditions in a grid much larger than the refuge width, starting from the initial condition corresponding to the homogeneous solution plus small random fluctuations, namely, $u(x,t=0)\simeq u_0+\xi(x)$, being $\xi$ an uncorrelated uniformly distributed variable  of amplitude much smaller than $u_0=r_{in}$. 
In numerical simulations, the refuge spans the interval 
$(-L/2,L/2)$, with fixed size $L=10$. 
The width of the rectangular kernel $\gamma$ is also fixed  ($w=1$).
The numerical results are complemented by analytical considerations based on linear stability analysis.

\subsection{Survival}
\label{sec:survival}

The  critical size of a refuge, for population survival, is known to depend on growth and diffusion coefficients, both inside and outside the refuge, when discontinuous binary forms of $r(x)$ and $D(x)$ are considered~\cite{maciel2013individual,colombo2016intermittent,colombo2018refuge, ludwig79,skellam91,kenkre03}. 
Succinctly, the critical size is derived assuming small $u$,  hence solving Eq.~(\ref{eq:FKPP}) in each domain after discarding nonlinearities, and coupling the solutions through continuity conditions. As the diffusivity is a continuous function in all the studied cases cases, then we impose continuity for $u$ and its derivative at the interfaces. 
Hence, the critical threshold for survival is   approximately   (see for instance \cite{colombo2016intermittent})  

 \begin{equation}
    L_c \simeq 2\sqrt{\frac{D_0}{r_{in}}} \arctan\left( \sqrt{\frac{-r_{out}}{r_{in}}} \right) \,,
    \label{eq:critical_sizeMain}
\end{equation} 
where $D_0$ is the diffusivity at the interface for low density. 
 Since we are interested in studying the patterns that can appear in these systems, we will work with values of the parameters far beyond the critical values for survival, i.e., $L\gg L_c$, to warrant that the population does not go extinct.

\subsection{Linear stability analysis}
\label{sec:stability}

For  small perturbations around the homogeneous state $u_0$, we can  substitute $u(x,t)\simeq u_0 + \epsilon(x,t) = u_0 +  \varepsilon \exp(ikx +\lambda t)$, with $\varepsilon \ll u_0$, into Eq.~(\ref{eq:FKPP}), obtaining the linearized form 
\begin{eqnarray} \label{eq:nl}
    \partial_t \epsilon & = &    
  \bar{D}  \partial_{xx} \epsilon -r \,\gamma \star \epsilon +\mathcal{O}(\epsilon^2) \simeq \lambda(k) \epsilon ,
\end{eqnarray}
where  $\bar{D}$ is the diffusion constant in the region of interest, that in the state-dependent cases becomes $\bar{D} =D(u_0)$.
For the rectangular kernel $\gamma(x)=\Theta(w-|x|)/(2w)$, Eq.~(\ref{eq:nl}) gives
\begin{equation} \label{eq:lambda}
     \lambda(k)= - \bar{D} \,k^2- r \frac{\sin{w k}}{w k}\,,
\end{equation}
which is the growth rate of mode $k$ that  can also be obtained by Fourier transforming Eq~(\ref{eq:nl}). 
Since at first order in this approximate approach the diffusivity is nearly constant, let us review the picture known for uniform diffusivity with periodic boundary conditions. 
If $\lambda(k)>0$ for some $k>0$,  the uniform density state is destabilized by a perturbation and  wrinkles develop. 
The dominant mode with wavelength $\Lambda^*=2\pi/k^*$, is given by the wavenumber
  $k^*$ which maximizes $\lambda(k)$. 
For Eq.~(\ref{eq:lambda}),   the first (global) maximum is found at $k^*\simeq 1.43\pi/w$, and imposing 
$\lambda(k^*)>0$,  one obtains the instability condition  ~\cite{kenkre03,colombo2012nonlinear} 
\begin{equation} \label{eq:cond1}
    \bar{D} \lesssim (1.43 \pi)^{-3} r w^2\simeq 0.011 r w^2 \equiv D_{c1}\,,
\end{equation}
when $r>0$;  otherwise, no \new{spatial} oscillations arise.  \new{Let us highlight that all oscillations observed in the studied system are not temporal but spatial.}

The condition (\ref{eq:cond1}) is \new{necessary} to produce patterns in a homogeneous landscape, with constant $r$. 
However, in a landscape with heterogeneous growth rate,  such as in the case of the refuge with $r(x)$ given by Eq.~(\ref{eq:rx}), and constant diffusivity $\bar{D}$, the uniform density state $u_0$ can be destabilized even for $\lambda(k)<0$ and \new{(spatially)} damped modes emerge from the complex roots of $\lambda(k)=0$~\cite{dornelas2021landscape}.   
In fact, linearization of Eq.~(\ref{eq:FKPP}), at first order in 
$\varepsilon$ and in $A=r_{in}-r_{out}$, leads to 
\begin{equation} \label{eq:eps}
    \tilde\varepsilon(k) = u_0\frac{\tilde{\psi}(k)}{-\lambda(k)}\,, 
\end{equation}
where ``\(\sim \)'' indicates Fourier transform and  $\psi(x)=A\Theta(|x|-L/2)$. 
\new{To obtain the induced perturbation in real space, we must compute the inverse Fourier transform of Eq.~(\ref{eq:eps}). 
By applying the residue theorem to this integration, it becomes clear that the perturbation takes the form 
\begin{align}\label{eq:sol_poles}
    \varepsilon(x) = \sum_{j} C_j\, e^{(i k_{R,j}   -  k_{I,j} x},
\end{align}
where, the oscillation parameters $k_{R,j}$ (wavenumber) and  $k_{I,j}$ (inverse of the decay-length) are the absolute values of the real and imaginary parts of the $j$-th zero of $\lambda(k)$, and the constant coefficients $C_j$ depend on the perturbation $\tilde{\Psi}$, which is assumed to be non-periodic (i.e., it does not add any characteristic mode by itself).
These roots can be obtained numerically. }

\new{The dominant mode (with lower damping) is given by the root $k=k_R\pm ik_I$ of $\lambda(k)$},  with the smallest imaginary part (in absolute value).  Its imaginary part gives the damping rate while the real part gives the associated wavelength $\Lambda=2\pi/|k_R|$ of the pattern. 
Therefore, depending on the real and imaginary parts of the relevant root, we can observe:

\noindent
(I) sustained \new{spatial} oscillations (when the imaginary part is zero),  which corresponds to condition (\ref{eq:cond1});  

\noindent
(II)  damped \new{spatial} oscillations (when real and imaginary parts are non-null);  

\noindent
(III) absence of \new{spatial} oscillations (when the real part is null). 

These roots can be obtained numerically, but for the selected rectangular kernel $\gamma$ it is possible to estimate the critical values analytically. 
A mathematical estimate  can be obtained from the Taylor expansion of  $\lambda(k)$ in Eq.~(\ref{eq:lambda}),  by finding the roots of the truncated series at a given order, which allows one to obtain an explicit relation between the parameters for the condition at which the real part of the relevant root becomes non null, yielding   
$\bar{D}/(r w^2)\lesssim  1/6+\sqrt{1/30} \simeq 0.35$  at fifth order, which can be exactly solved, and including  higher order terms of the Taylor expansion, we numerically arrive at 
\begin{equation} \label{eq:cond2}
    \bar{D} \lesssim 0.37 r w^2\equiv D_{c2}.   
\end{equation} 
This is the condition for the appearance of oscillations induced by growth-rate heterogeneity, for constant diffusion coefficient $\bar{D}$. For heterogeneous diffusivity, inside the refuge, we have $\bar{D} \simeq D_{in}$ for the binary case and $\bar{D}\simeq D(u_0)$ for the state-dependent profiles.

\section{Results for state-dependent diffusivity}
\label{sec:stateD}

In this section, we focus on diffusivity coefficients that are ruled by the density. 
Since the diffusivity depends on the population density, both evolve  in time concomitantly, in a self-consistent way. 

\subsection{Enhanced \new{diffusion due} to sparsity:    $D_1(u)=d\exp(-u/\sigma)$}
\label{sec:sparsity}
 
\begin{figure}[b!]
    \centering
    \includegraphics[width=0.49\columnwidth]{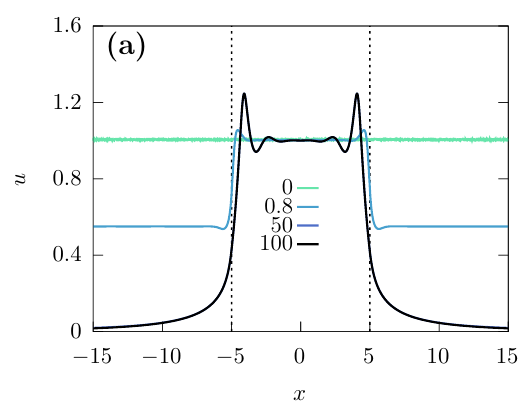} \\
    \includegraphics[width=0.49\columnwidth]{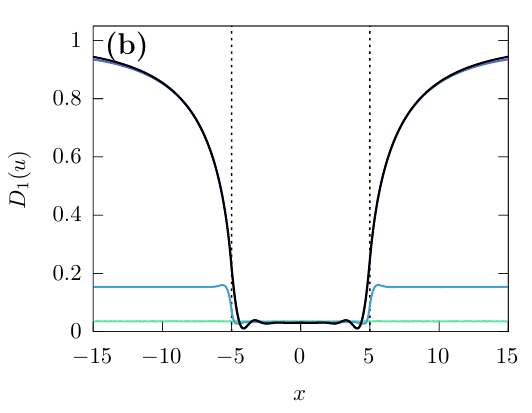} 
    \includegraphics[width=0.49\columnwidth]{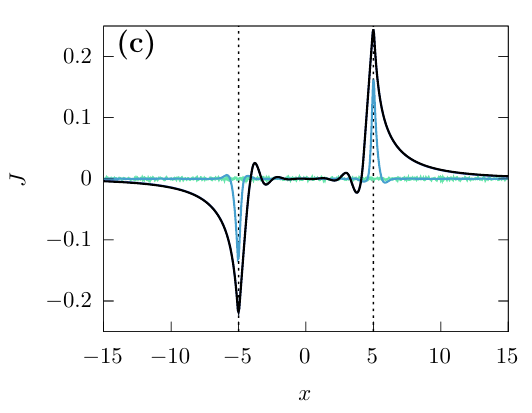}  
    \caption{Time evolution of population density (a), the density-dependent diffusion coefficient $D_1(u)$ (b), and flux $J=-D_1(u)\partial_x u$ (c), for times $t$ indicated in the legend, fixing $\sigma=0.3$, $d=1$, $r_{in}=1$, for   $r_{out}=-0.02$.  
    }
    \label{fig:diffusion_Time}
\end{figure}

Fig.~\ref{fig:diffusion_Time} provides an illustrative example of the time evolution of the system in the presence of a refuge,
starting from a noisy uniform density around $u_0$, until reaching a long-time profile (which occurs for $t\gtrsim 100$ in the case of the figure).

Initially (at $t=0$),  the density, hence the diffusivity,  are nearly constant, 
satisfying $ D_{c1}< d\exp(-u_0/\sigma)\simeq 0.036<D_{c2}$ (according to Eqs.~(\ref{eq:cond1})-(\ref{eq:cond2}), having used $u_0=r=w=1$, $d=1$ and $\sigma=0.3$). 
In a first regime ($t\lesssim 1$), the two-level profile of the growth rate $r(x)$ induces essentially a two-level population density,  which in turn molds the diffusivity profile. 
As soon as Eq.~(\ref{eq:cond2}) holds, steady oscillations emerge, with the characteristic wavelength $\Lambda^*$ predicted by the linear analysis. 
 
In Fig.~\ref{fig:diffusion_Time}, we used $r_{out}=-0.02$, implying a weakly hostile environment. In the more lethal case $r_{out}\to-\infty$, the external populations will go extinct. 
Inside the refuge, the higher peaks of the density profile shift towards the center when $|r_{out}|$ grows (not shown), but the qualitative features remain essentially the same. Then we will choose 
$r_{out}=-0.02$ so that the variations outside the refuge are amplified.   

\begin{figure}[t!]
    \centering
    \includegraphics[width=0.49\columnwidth]{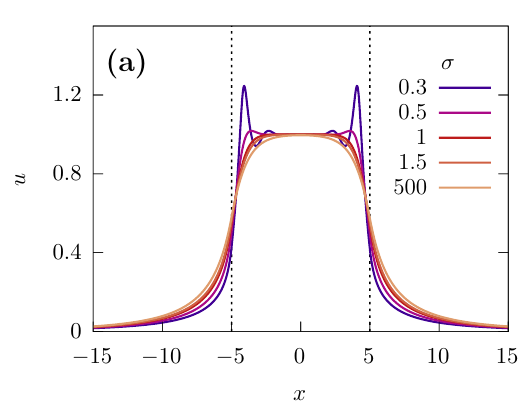}
     \includegraphics[width=0.49\columnwidth]{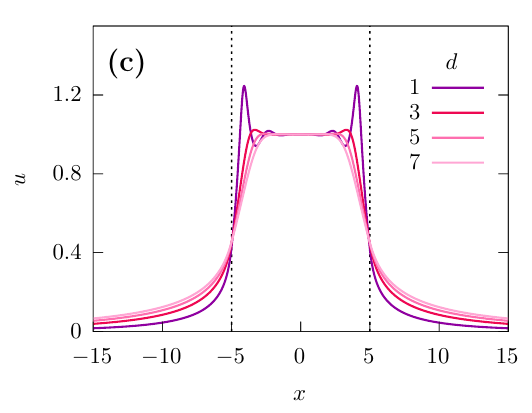}
 \includegraphics[width=0.49\columnwidth]{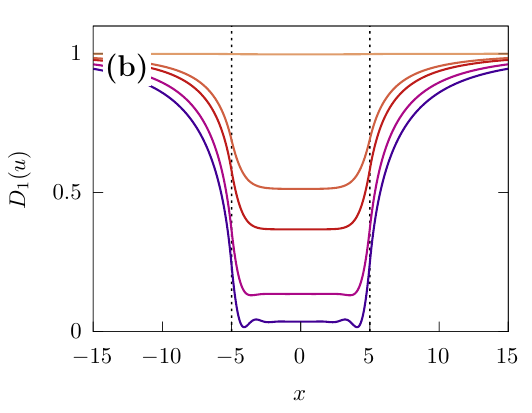} 
  \includegraphics[width=0.49\columnwidth]{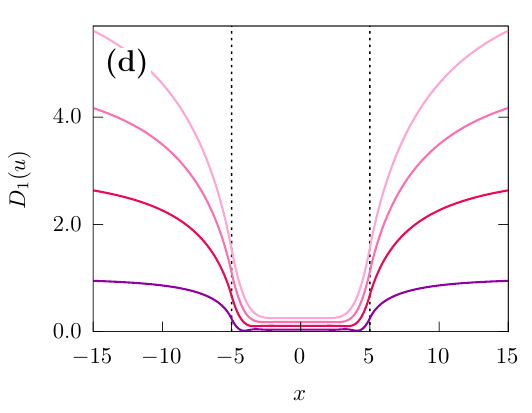} 
    \caption{Stationary profiles $u$, and corresponding $D_1(u)=d \exp(-u/\sigma)$,   (a)-(b) with different values of $\sigma$ indicated in legend, keeping $d=1$; 
    (c)-(d), with different values of $d$, 
    setting  $\sigma=0.3$. Fixed parameters are $r_{in}=1$, $r_{out}=-0.02$.
    }
    \label{fig:Du}
\end{figure}

With regards to the parameters that define $D_1(u)$, let us remark that, while the pre-factor $d$ regulates the maximum of diffusivity, $\sigma$ controls how sensitive to density is the response of the diffusion coefficient. 
Let us analyze their influence on the stationary profiles, looking at Fig.~\ref{fig:Du}.

 Approximately, 
 $D_{in}\simeq d \exp(-u_0/\sigma)$ in the refuge and $D_{out}\simeq d$ outside far from the interface. 
 Therefore, $\sigma$ affects the internal diffusivity, producing a level that can impact pattern formation, as well as the external population close to the interface, affecting the persistence of the population in this region.  
For increasing $\sigma$,   when the internal diffusivity level exceeds the critical value $D_{c2}\simeq0.37$  (according to Eq.~(\ref{eq:cond2}), and recalling that we set $r=w=1$),  oscillations are spoiled, 
and the dependence on $u$ is flattened, producing homogeneous diffusivity.  
 In the opposite case of small $\sigma$, the diffusivity is more sensitive to the variations of density and results a density profile with sharp peaks near the interfaces. \new{For even smaller values of $\sigma$ or $d$ than those shown in the figure, the peaks grow without attaining a steady state (not shown).}

Moreover, from the viewpoint of the maintenance of the population level, 
a larger $\sigma$, which tends to homogenize the diffusivity, 
reduces the internal population, mainly the high crowding near the interface, but enhances the outer population, which despite the negative growth rate is fed by the outward fluxes. 

The pre-factor $d$ affects both $D_{in}$ and $D_{out}$, as can be seen in  
Fig.~\ref{fig:Du}(c)-(d), 
 influencing  the oscillations inside the refuge and the decay outside. 
In this case, a variation of $d$ tending to homogenize the diffusivity, which occurs at a low level, benefits the increase of the internal population to the detriment of the external one.

\begin{figure}[h]
\includegraphics[width=0.7\columnwidth]{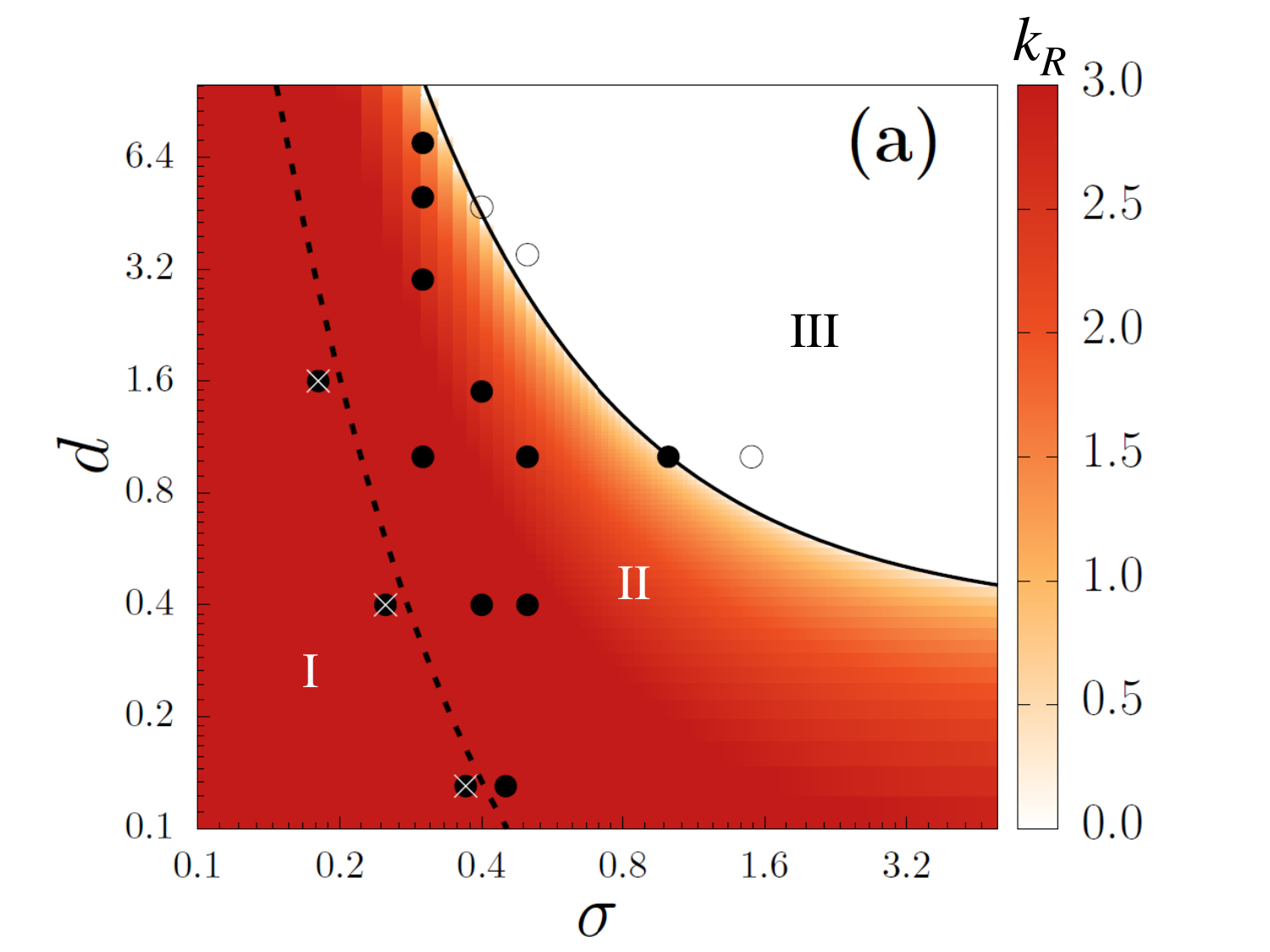}
\includegraphics[width=0.7\columnwidth]{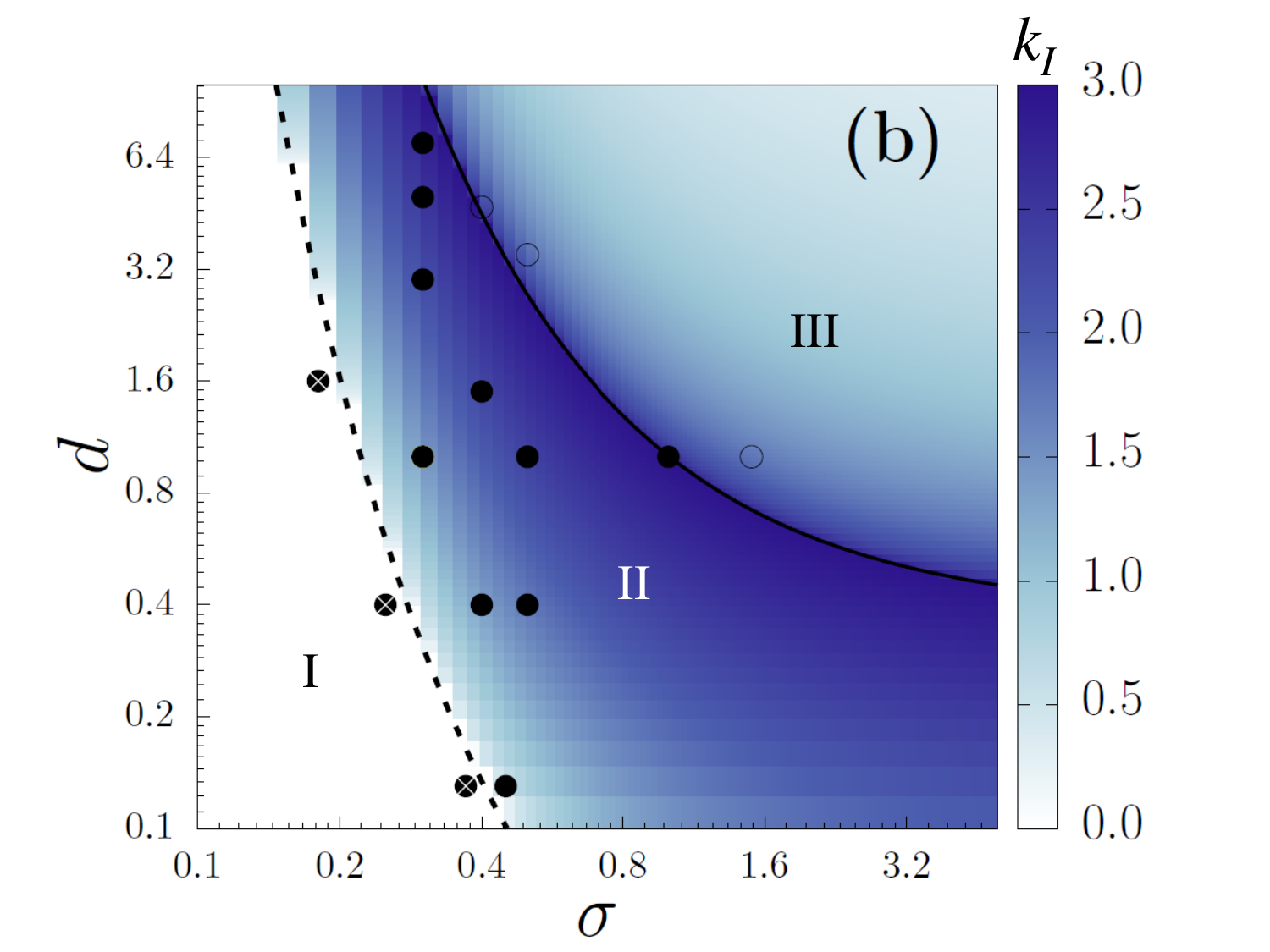}
     \caption{Phase diagrams in the plane $\sigma-d$ of $D_1(u)$.   
 The regions I-III correspond to those described in Sec.~\ref{sec:methods}, delimited 
 by dashed and full lines, given by 
Eqs.~(\ref{eq:cond1})  and (\ref{eq:cond2}), respectively, using $\bar{D}=D_1(u_0)$.
The heat plots represent the real $k_R$ (a) and imaginary $k_I$ (b) parts of the dominant root of $\lambda(k)$.   
The symbols  correspond to the type of density profile observed in the  numerical long-time solution of Eq.~(\ref{eq:FKPP}): 
damped oscillations (filled),  
uniform (hollow), and for the crossed circles a steady state is not observed. 
Fixed parameters are $r_{in}=1$, $r_{out}=-0.02$. 
     }
     \label{fig:Diagrams}
\end{figure}

The effects of both parameters of $D_1(u)$ on the oscillations inside the refuge are summarized in Fig.~\ref{fig:Diagrams},  where we depict phase diagrams in the plane $\sigma-d$, identifying the different structures that can emerge. 

The dashed and full lines correspond to the conditions given by 
Eqs.~(\ref{eq:cond1})  and (\ref{eq:cond2}), respectively, using 
$\bar{D}=D(u_0)$, which delimit the different phases I-III (see Sec.~\ref{sec:methods}). 
Despite the oscillations of the density, the approximation of a constant diffusivity 
$D(u_0)$ in the internal region gives a prediction in reasonable agreement with the results from simulations. 
We verified that pattern wavelength agrees with the prediction $\Lambda=2\pi/|k_R|$. 
Nonetheless,
in the region where the imaginary part vanishes and sustained oscillations are expected,
what actually happens is that a steady state is not attained (crossed symbols).  
 A high peak gives rise to a locally low diffusion coefficient, then the 
 fluxes outwards from the peak are not able to balance the growth of the 
 population and this makes the peak even higher, an effect which is reinforced the higher the peak. 

Notice that such feedback does not take place for density-independent diffusivity, in which case a steady state is always achieved.     
We also highlight that for $D_1(u)$, when stationary oscillations are formed in the refuge, they are strongly damped towards the center.
Moreover, the population outside the refuge 
can attain a moderate level at the interface and decay  slowly with the distance to the interface.

\subsection{Enhanced \new{diffusion due} to overcrowding: $D_2(u)=d[1-\exp(-u/\sigma)]$}

\begin{figure}[b!]
    \centering
     \includegraphics[width=0.49\columnwidth]{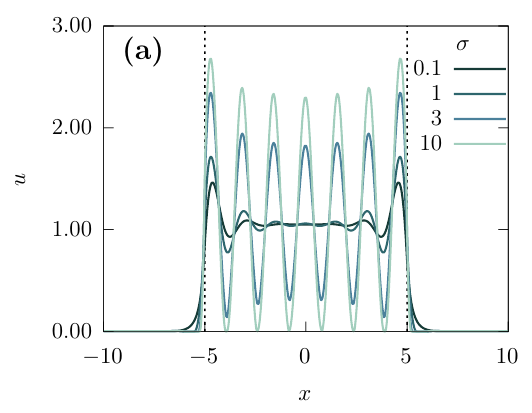} \\
     \includegraphics[width=0.49\columnwidth]{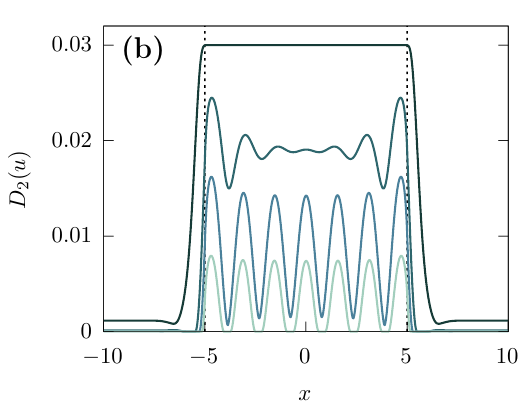} \includegraphics[width=0.49\columnwidth]{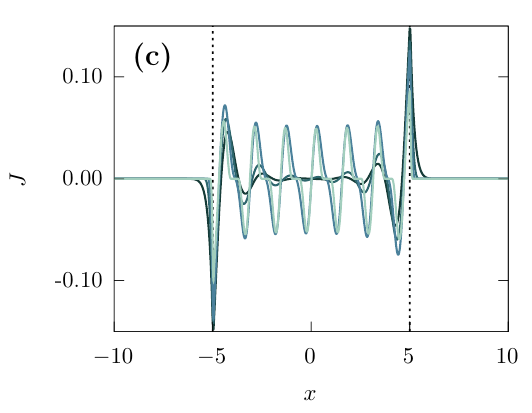}
    \caption{Stationary profiles of $u$, $D(u)$ and flux $J$, generated by $D(u)=D_2(u)=d[1-\exp(-u/\sigma)]$,  for different values of the decay parameter $\sigma$, with fixed $d=0.03$, $r_{in}=1$, $r_{out}=-0.02$.
    }
    \label{fig:overcrowding}
\end{figure}

In this section we discuss the effects of the diffusivity $D_2(u)$ defined in Eq.~(\ref{eq:Du2}), 
which  presents a higher level in response to overcrowding.  
For large enough $d$, the high internal diffusivity level forbids pattern formation, then we considered low values of $d$ only. 
Typical plots of population density and diffusivity  profiles are shown in Fig.~\ref{fig:overcrowding}.

Although the  diffusivity in the outer region becomes lower that in the refuge, oscillations are not formed due to the negative growth rate.
Inside the refuge, the diffusivity is nearly constant ($D_2 \simeq d$), when $\sigma \ll u_0$. As $\sigma$ increases, the average level of the diffusivity decreases, favoring oscillations, which tend to become sustained when the diffusivity profile becomes bounded by $D_{c1}$. 
Furthermore, due to the particular dependency on $u$, the diffusivity is higher at the peaks of $u$ and very small at the valleys, favoring a steady state with barely sustained oscillations, differently to what we observed in the opposite case of reaction to sparsity. 
Moreover, for sufficiently large $\sigma$, as the amplitude of the sustained oscillation grows, 
fragmentation  of the population inside the refuge can occur (with bumps separated by depopulated regions), a phenomenon  observed  for the space-dependent profile but not 
for $D_1(u)$.  \new{Note that $D_2(u)$ tends to zero when $u$ tends to zero, but similar features arise with a baseline different from zero (not shown).}

With regard to the distribution of the population outside the refuge, notice in Fig.~\ref{fig:overcrowding}, the low level at the interface and fast decay, even when $r_{out}$ is only slightly negative. This is another difference with the heterogeneous diffusivity $D_1(u)$, which allows to a significant external population near the interface. 
Succinctly, in the case of $D_1$ which reflects a reaction to sparsity, the flux depopulate the peaks inside the refuge and produces a smooth external decay. Differently, the reaction to overcrowding represented by $D_2(x)$ helps to populate the peaks and to produce a faster decay from the interface.

\begin{figure}[b!]
    \centering
\includegraphics[width=0.49\columnwidth]{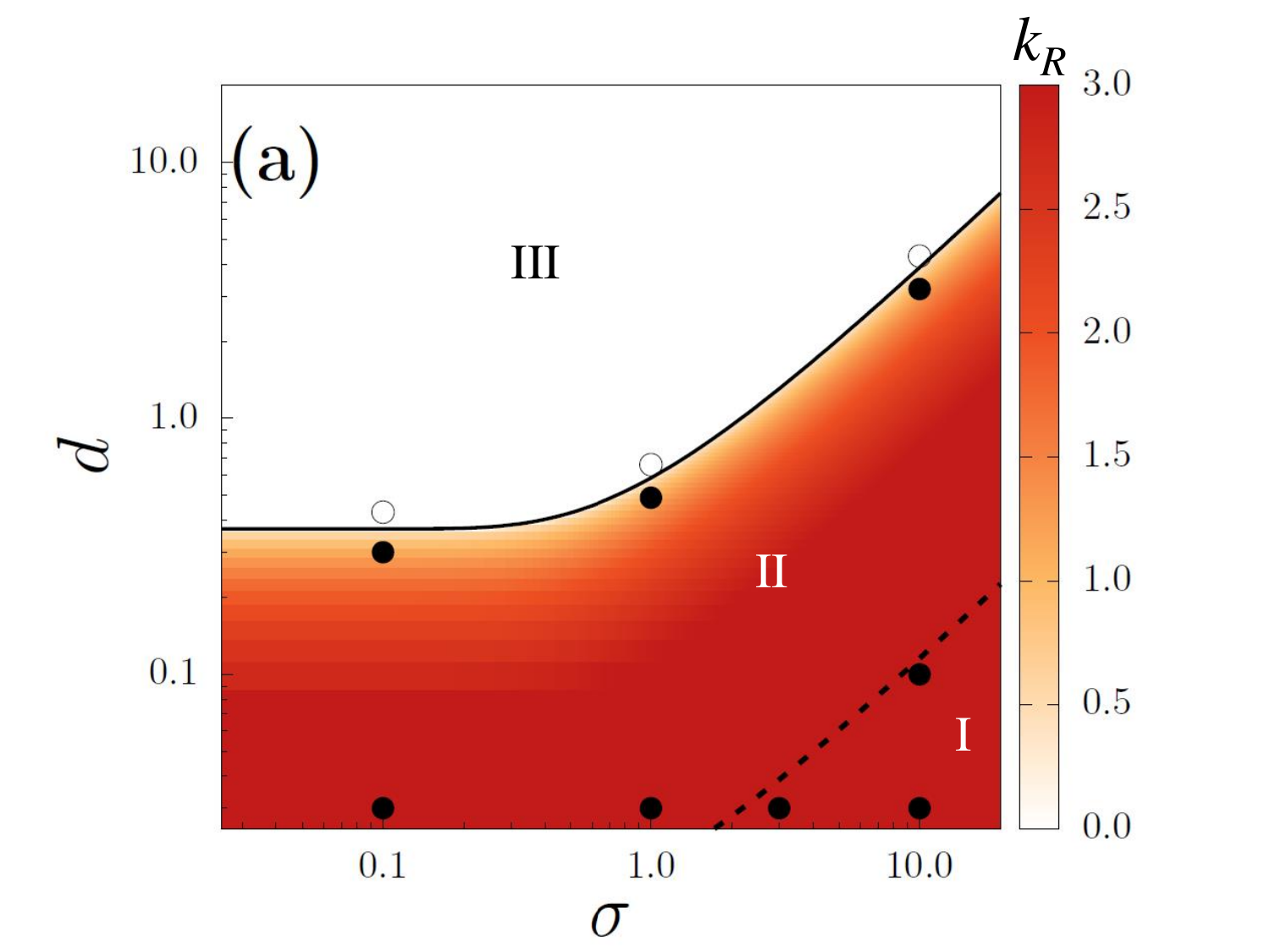}
\includegraphics[width=0.49\columnwidth]{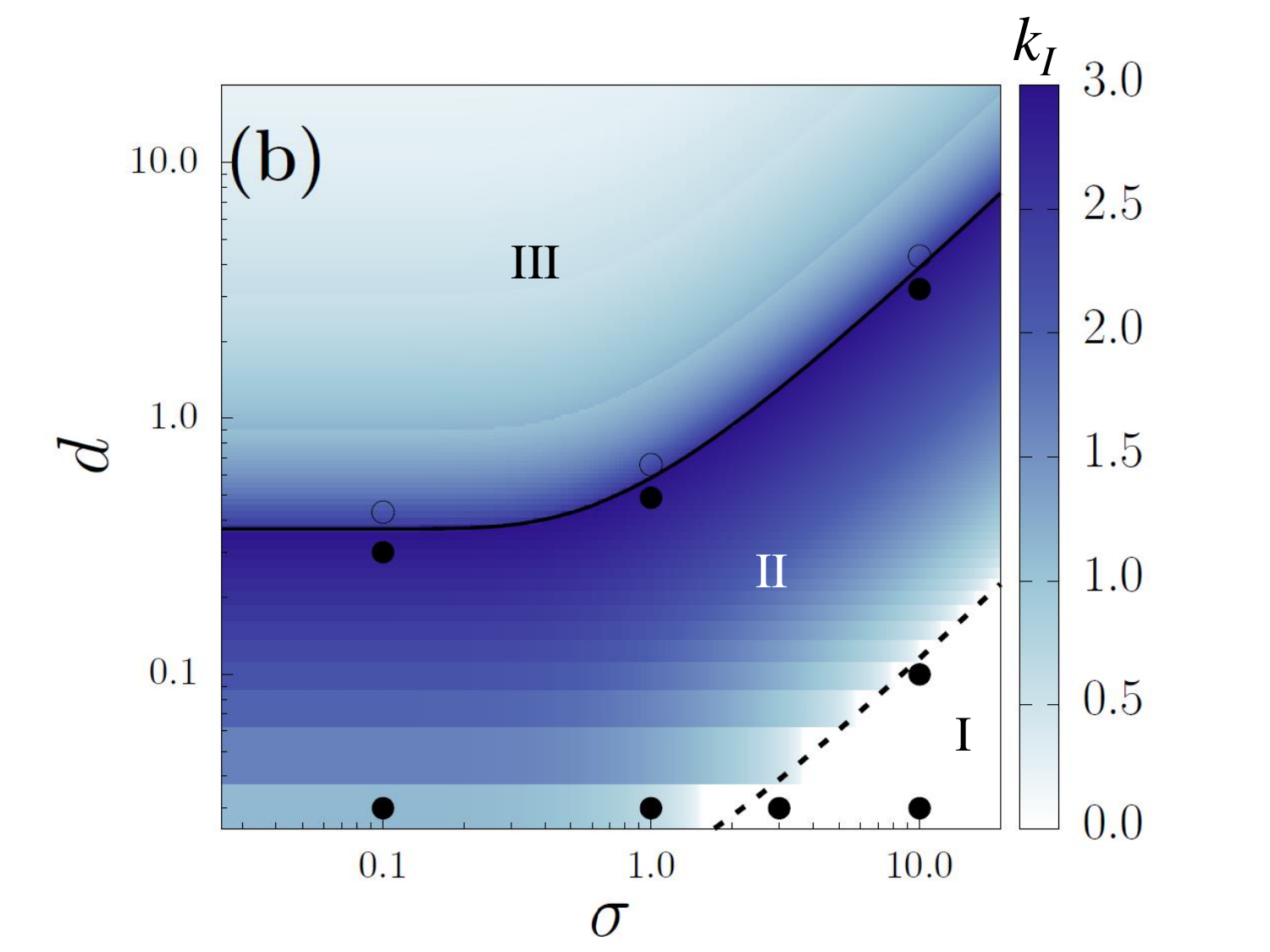}
    \caption{
    Phase diagrams in the plane $\sigma-d$ of $D_2(u)$.  
    The regions I-III correspond to those described in Sec.~\ref{sec:methods}, delimited 
 by dashed and full lines, given by 
Eqs.~(\ref{eq:cond1})  and (\ref{eq:cond2}), respectively, using $\bar{D}=D_2(u_0)$.
As in Fig.~\ref{fig:Diagrams}, the heat plots represent the real $k_R$ (a) and imaginary $k_I$ (b) parts of the dominant root of $\lambda(k)$.   
The symbols  correspond to the type of density profile observed in the  numerical long-time solution of Eq.~(\ref{eq:FKPP}): 
damped oscillations (filled),  
uniform (hollow). 
 Fixed parameters $r_{in}=1$, $r_{out}=-0.02$. 
 }
    \label{fig:Diagrams2}
\end{figure}

 The effect of the parameters is summarized in the phase diagrams shown in Fig.~\ref{fig:Diagrams2}. In general there is a good agreement between the classification of steady states (symbols) and the predicted regions. 
However, where the linear approximation predicts sustained oscillations (white region in panel (b)), actually, a slight damping occurs, as can be seen in Fig.~\ref{fig:overcrowding} for the case $\sigma =10$  and $d=0.03$. 
Also note that the theoretical prediction based on the approximation 
$u\sim u_0$ for the inner population can fail when large amplitude   oscillations develop.

\section{Results for space-dependent diffusivity} \label{sec:spaceD}

Let us consider the case where the diffusion coefficient is different inside and outside the refuge,  independently of the population level. 
Instead of imposing continuity conditions~\cite{maciel2013individual}, we opted for a smoothed form $\Theta_s(x)$ of the Heaviside function, in Eq.~(\ref{eq:Dx}), connecting the two levels by a narrow but finite interface of width $s \ll w$,  by defining

\begin{equation}
\Theta_s(x)= \frac{\tanh\left(  \frac{x-L/2}{s}\right)+\tanh\left( \frac{x+L/2}{s}\right)}{2 \tanh(\frac{L}{  2 s} )},
 \label{eq:thetas}
\end{equation}
which tends to the Heaviside step function when $s \to 0$.
In analogous way, we might smooth the growth rate profile, but since its derivatives do not enter the evolution equation, this smoothing is not significant in this case.

\begin{figure}[b!] 
    \centering
\includegraphics[width=0.49\columnwidth]{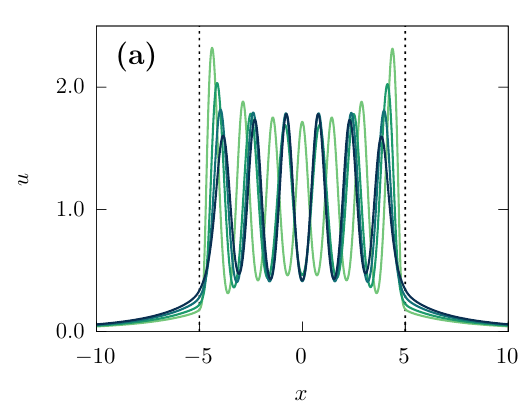} 
\includegraphics[width=0.49\columnwidth]{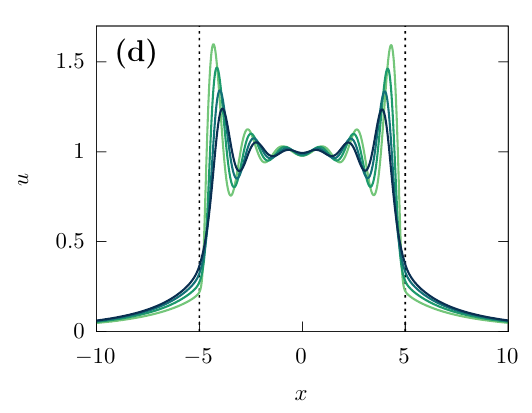}  \includegraphics[width=0.49\columnwidth]{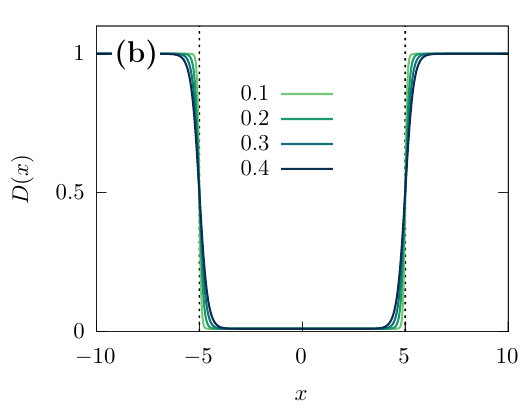}
\includegraphics[width=0.49\columnwidth]{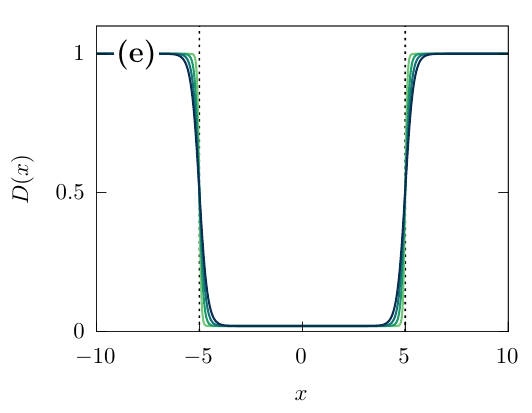} 
\includegraphics[width=0.49\columnwidth]{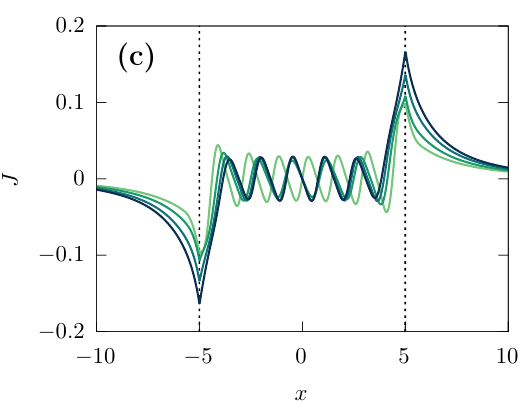} 
\includegraphics[width=0.49\columnwidth]{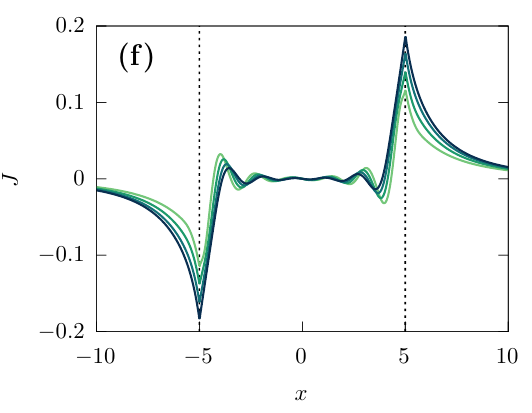}
    \caption{Stationary profiles of $u(x)$, $D(x)$ and $J$, under space-dependent diffusivity in Eq.~(\ref{eq:Dx}), for a refuge of size $L=10$ signaled by vertical dotted lines, with different values of the interface width $s$ indicated in the legend,  
    for $D_{in}=0.01$ in (a)-(c) and for  $D_{in}=0.02$ in (d)-(f). Fixed parameters are $D_{out}=1$, $r_{in}=1$, $r_{out}=-0.02$.
    }
\label{fig:Dvsx}
\end{figure}

Typical stationary profiles of $u(x)$,  $D(x)$  and the flux $J=-D(x)\partial_x u$ are shown in Fig.~\ref{fig:Dvsx}, changing the interface width $s$, for two different values of $D_{in}$.  
Increasing $s$ affects the effective width of the refuge and the wavelength of the oscillations is stretched, diminishing the number of peaks. 
When $s$ decreases, the intensity of the flux at $\pm L$ decreases, allowing the growth of the peak closer to the interface, while outside the refuge, the density becomes lower near the interface. 
Then in the limit $s\to 0$, a jump in the first derivative of the density is expected, such that   the current $J=-D(x)\partial_x u$ remains continuous at the interfaces. 
The fluxes are maximal at the interfaces, pointing outwards. 
Inside the refuge, the fluxes tend to depopulate the peaks (as observed for $D_1$), but the logistic growth balances the fluxes and a steady state is attained.   
Independently of the  width $s$, for $D_{in}=0.01$ (a)-(c), Eq.~(\ref{eq:cond1}) approximately holds and oscillations are barely  sustained, except for the external peak, while they  are strongly damped for $D_{in}=0.02$ (d)-(f).   
Since small values of $s$ produce qualitatively similar results, we fix $s=0.2$.

\begin{figure}[b!]
    \centering
\includegraphics[width=0.49\columnwidth]{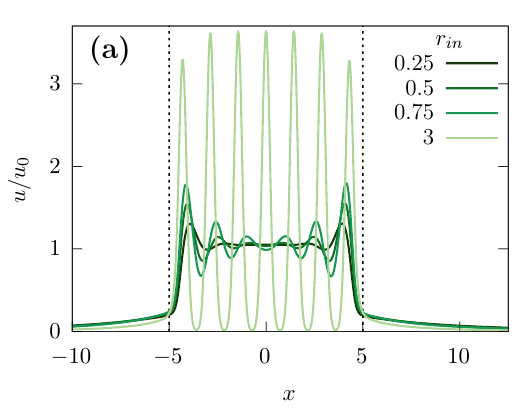}
\includegraphics[width=0.49\columnwidth]{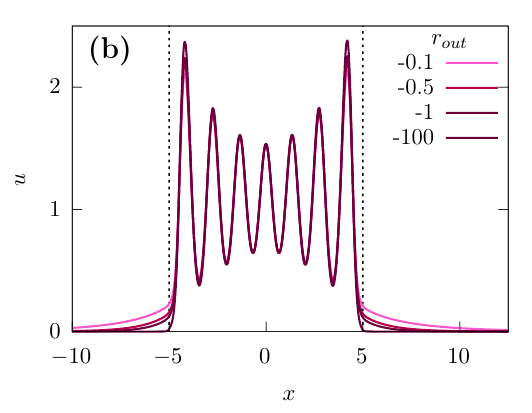} \\
\includegraphics[width=0.49\columnwidth]{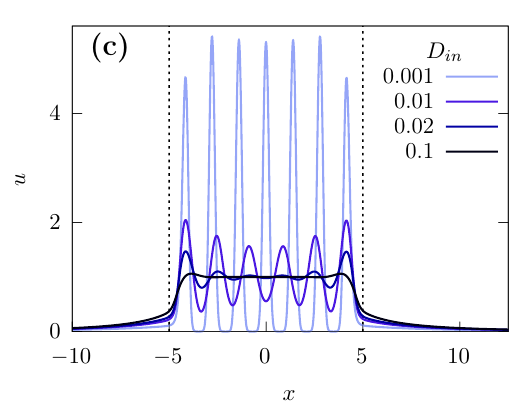} 
\includegraphics[width=0.49\columnwidth]{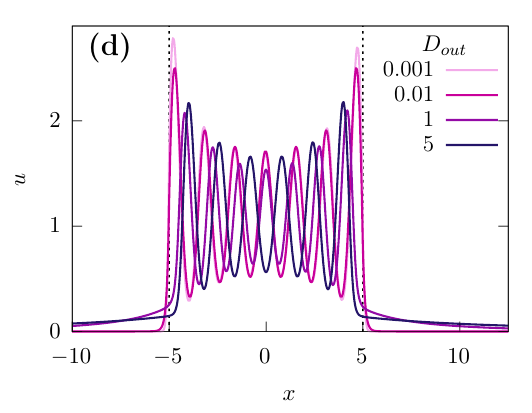}
    \caption{Stationary profiles of the population density in a refuge, for  $s=0.2$,  and varying the parameters indicated in each panel, fixing 
    (a) $r_{out}=-0.02$,  $D_{in}=0.01$, $D_{out}=1$,
    (b) $r_{in}=1$, $D_{in}=0.01$, $D_{out}=1$, 
    (c) $r_{in}=1$, $r_{out}=-0.02$, $D_{out}=1$, 
    (d) $r_{in}=1$, $r_{out}=-0.02$, $D_{in}=0.01$.
    }
   
    \label{fig:all}
\end{figure}

In Fig.~\ref{fig:all} we show the effects of 
varying   $D_{in}$ and  $D_{out}$, as well as $r_{in}$ and $r_{out}$. 
In (a) we observe the expressive reduction of damping and increase of the oscillation amplitude   associated to the increase of the inner growth rate $r_{in}$. 
A similar effect is produced by decreasing $D_{in}$, in (c), as predicted by the linear analysis.
Furthermore, notice that the profiles can become fragmented inside the refuge, with regions of  nearly null density between bumps.  

Values of $r_{out}<0$, varying in a wide range, have little effect on the internal population, 
but the external population decays faster from the interface with increasing $|r_{out}|$, without fluctuations, vanishing in the limit of fully lethal conditions $r_{out}\to-\infty$. 
Increasing $D_{out}$ reduces the population in the refuge close to the interface, changing the wavenumber of the oscillations, and makes the external population decay more slowly with distance. 
 In fact, in the external region, the population density decays nearly exponentially from the interface, as
$ u(x)\sim \exp\left(-\sqrt{\frac{r_{out}}{ D_{out}}} [x-L/2] \right)$, as results from the small $u$ approximate solution of Eq.~(\ref{eq:FKPP}). 
When $D_{out}<D_{in}$, the population vanishes already near the interface.

\begin{figure}[b]
    \centering
\includegraphics[width=0.49\columnwidth]{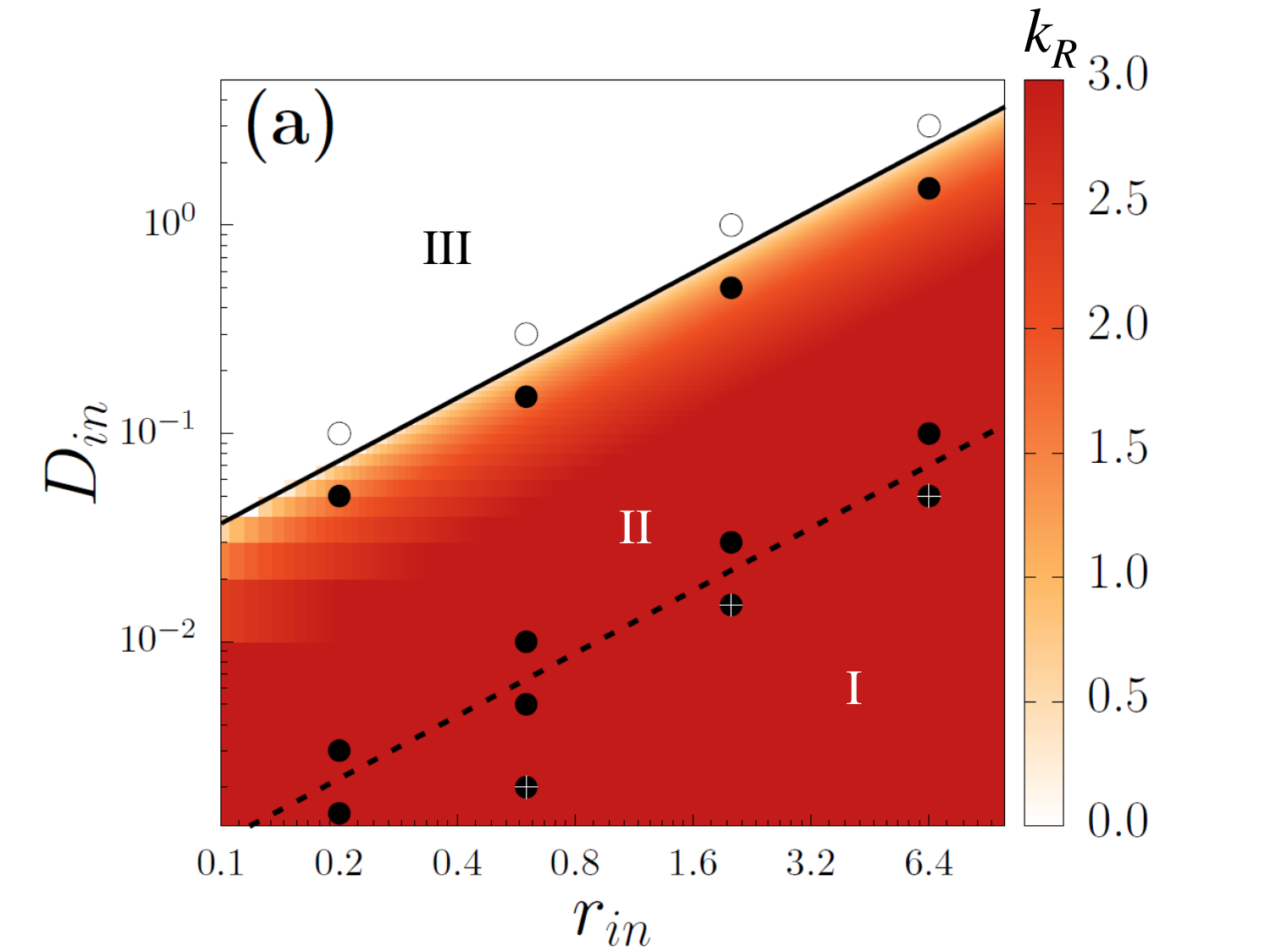}
\includegraphics[width=0.49\columnwidth]{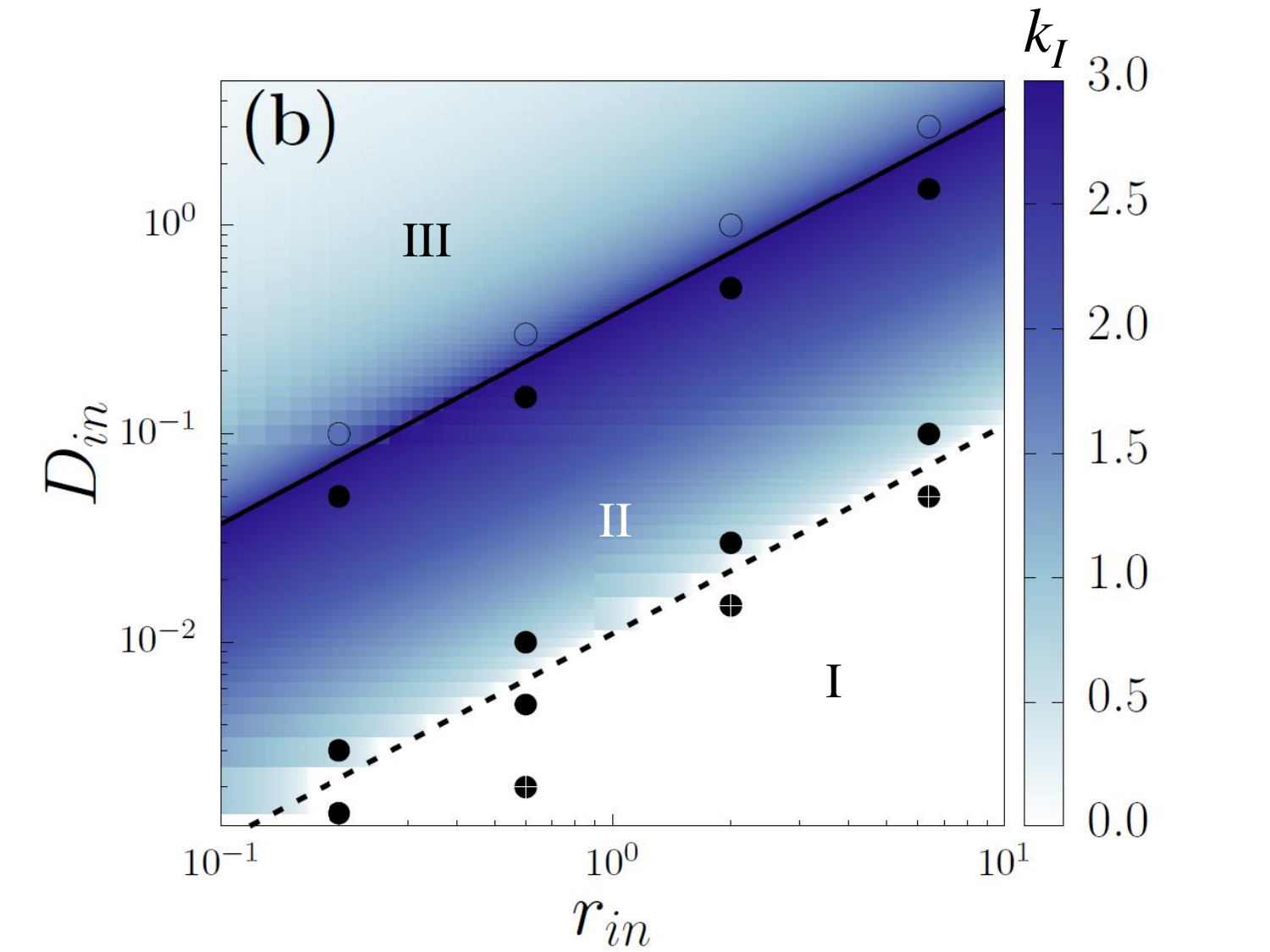}
    \caption{
    Phase diagrams in the plane $D_{in}-r_{in}$ for the binary profile $D(x)$.
     The regions I-III correspond to those described in Sec.~\ref{sec:methods}, 
  delimited by dashed and solid lines given by Eqs. (\ref{eq:cond1}) and (\ref{eq:cond2}), respectively, setting $\bar{D}=D_{in}$.      
     The heat plots represent the real $k_R$ (a) and imaginary $k_I$ (b) parts of the dominant root of $\lambda(k)$. 
 The symbols  correspond to the type of density profile observed in the  numerical long-time solution of Eq.~(\ref{eq:FKPP}): 
sustained oscillations (plus-circles), 
damped oscillations (filled), and
uniform state (hollow), in agreement with regions I-III.  
 Fixed parameters are $D_{out}=1$, $r_{out}=-0.02$. 
 }
    \label{fig:Diaginout}
\end{figure}

We summarize the effects produced on pattern formation by the two more relevant parameters 
$D_{in}$ and $r_{in}$, in the phase diagrams of Fig.~\ref{fig:Diaginout}.
The predictions of the linear analysis in Sec.~\ref{sec:methods} (for $\bar{D}=D_{in}$ and $r=r_{in}$)  are in accord with the kind of solutions (symbols) obtained from numerical integration of Eq.~(\ref{eq:FKPP}). 
The dashed and full lines correspond to the conditions given by 
Eqs.~(\ref{eq:cond1})  and (\ref{eq:cond2}), respectively, which delimit the different phases. 
The color plots represent the real (a) and imaginary (b) parts of the dominant eigenvalue (minimum positive imaginary part, see Sec.~\ref{sec:methods}), obtained numerically. 
Recall that, in (a) the white region (zero real part) means no oscillations, in (b) the white region (zero imaginary part) predicts non-damped oscillations, otherwise damped oscillations with multiple peaks emerge within the refuge. 
Filled and open symbols correspond to the classification of steady profiles obtained by numerical integration of the FKPP Eq.~(\ref{eq:FKPP}), for which oscillations are observed, or not, respectively, in good accordance to the regions predicted by the theoretical critical curves.

For the ranges of the parameters in this figure, a steady state is always attained, and sustained oscillations are observed near the expected frontier (dashed line),

\section{Final remarks}

We have discussed the impact of heterogeneous diffusion 
 on the spatial organization of a population 
 in a refuge immersed within a hostile environment. 
We considered two  forms of state-dependent diffusivity $D(u)$ that mimic opposite reactions to the density, and also space-dependent diffusivity $D(x)$ where the heterogeneity is anchored on the quality of the environment, independently of the population.

In all the analyzed cases, 
when the average diffusivity level is sufficiently low inside the refuge, for the other parameters fixed, internal patterns emerge. 
Regardless of the details of the diffusivity profile inside the refuge, 
its average level  allows a good prediction  of the frontiers for patterning through the linear stability analysis. 
This indicates a certain robustness of the process of pattern formation against heterogeneities in the diffusivity profile.  

Nevertheless, the type of heterogeneity has effects on the shape of the patterns, not captured by the linear analysis.  
  In the case of overcrowding reaction, it is possible to have more uniform peaks and fragmentation can occur, similar to what happens with the two-level profile.
In the case of mobility enhanced by sparsity through $D_1$, the peaks rise higher the closer to the interface and a steady state may be unattainable due to progressive condensation in these peaks feedback by the dependence on density.

Outside the refuge, the population decays from the interface more slowly with higher levels of the external diffusivity and within the setting represented by $D_1$,   where  the diffusivity is low for high concentrations. The possibility of distinct diffusivity levels inside and outside could explain profiles observed experimentally, which present 
tails from the interfaces with moderate density of individuals even under very adverse conditions. 

Our numerical results and theoretical considerations, exploring how density-dependent diffusivity affects the organization of the population inside and outside the refuge, reveal  macroscopic signatures that could provide insights about hidden mechanisms. Moreover, this knowledge can be used to refine mathematical models. Within the particular context of the FKPP equation, in previous attempts to explain  the distribution of populations of   bacteria in  an environment subject to UV light with a refuge~\cite{perry2005experimental},  the diffusivity, measured for low bacterial density, is too high to allow destabilization of the homogeneous state empirically observed.  But if the mobility is affected by density, being lower in the crowded refuge, or distinct under favorable or  hostile conditions, oscillations inside and  tails outside the refuge would be explained.  
In fact some of the profiles obtained in this work remind those observed in Perry's experiments, but a deeper study is necessary in this direction, and this work may represent a starting point.

{\bf Acknowledgements: }
We are grateful to Eduardo H.F. Colombo for insightful discussions.
We acknowledge partial financial support by the Coordenação de
Aperfeiçoamento de Pessoal de Nível Superior - Brazil (CAPES) -
Finance Code 001. C.A. also acknowledges partial support received from
 Conselho Nacional de Desenvolvimento Científico e Tecnológico (CNPq)-Brazil (311435/2020-3) and Fundação de Amparo à Pesquisa do
Estado de Rio de Janeiro (FAPERJ)-Brazil (CNE E-26/201.109/2021,E-26/204.130/2024).


\end{document}